\documentclass{article}
\usepackage{amsmath,graphicx,mlspconf}
\usepackage{float}

\usepackage{multirow}

\usepackage{stfloats}
\usepackage{pgfplots}
\usepackage{svg}  
\usetikzlibrary{plotmarks}
\usetikzlibrary{arrows.meta}
\usepgfplotslibrary{patchplots}
\pgfplotsset{compat=1.17}

\copyrightnotice{978-1-6654-8547-0/22/\$31.00
{\copyright}2022 European Union}

\toappear{2022 IEEE International Workshop on Machine Learning for Signal Processing, Agu.\ 22--25, 2022, Xi'an, China}

\title{Floor map reconstruction through radio sensing and learning by a large intelligent surface}
\usepackage{cite}
\usepackage{amsmath,amssymb,amsfonts}
\usepackage{algorithmic}
\usepackage{graphicx}
\usepackage{textcomp}
\usepackage{xcolor}
\def\BibTeX{{\rm B\kern-.05em{\sc i\kern-.025em b}\kern-.08em
    T\kern-.1667em\lower.7ex\hbox{E}\kern-.125emX}}

\usepackage[nowarn,acronyms,nonumberlist,nopostdot,nomain,nogroupskip]{glossaries}

\usepackage{caption}
\usepackage{subcaption}

\newacronym{lidar}{LIDAR}{Light Detection and Ranging}
\newacronym{rf}{RF}{Radio Frequency}
\newacronym{mimo}{MIMO}{Multiple-Input-Multiple-Output}
\newacronym{bs}{BS}{Base Station}
\newacronym{lis}{LIS}{Large Intelligent Surface}
\newacronym{dgm}{DGM}{Deep Generative Model}
\newacronym{cgan}{cGANs}{conditional Generative Adversarial Networks}
\newacronym{vae}{VAE}{Variational Autoencoder}
\newacronym{mf}{MF}{Matched Filter}
\newacronym{gan}{GANs}{Generative Adversarial Networks}
\newacronym{nn}{NN}{Neural Network}
\newacronym{snr}{SNR}{Signal-to-Noise Ratio}
\newacronym{psnr}{PSNR}{Peak Signal-to-Noise Ratio}
\newacronym{mse}{MSE}{Mean Squared Error}
\newacronym{dof}{DoF}{Degrees of Freedom}
\newacronym{ls}{LS}{Least Square}
\newacronym{unet}{UN}{U-Net}
\newacronym{ssim}{SSIM}{Structural Similarity Index Measure}

\usepackage{booktabs}
\usepackage{pgfplots}
\usepackage{svg}  
\usetikzlibrary{plotmarks}
\usetikzlibrary{arrows.meta}
\usepgfplotslibrary{patchplots}

\usepackage{svg}

\pgfplotsset{compat=1.17}

\begin{document}

\name{
    Cristian J. Vaca-Rubio$^{\star,  \mathsection}$, Roberto Pereira$^{\dagger,  \mathsection}$, Xavier Mestre$^{\dagger}$, David Gregoratti$^{\ddagger}$, \\ \textit{Zheng-Hua Tan}$^{\star}$, \textit{Elisabeth de Carvalho}$^{\star}$ and \textit{Petar Popovski}$^{\star}$ \thanks{$\mathsection$ Both authors contributed equally to this research. This work has been partially funded by the European Commission under the Windmill project (contract 813999) and the Spanish government under the Aristides project (RTI2018-099722-B-I00). This work has been submitted to IEEE for possible publication. Copyright may be transferred without notice, after which this version may no longer be accessible. 
    }}
\address{
    $^{\star}$ Department of Electronic Systems, Aalborg University, Aalborg, Denmark. \\ Email: \{cjvr, zt, edc, petarp@es.aau.dk\} \\
    $^{\dagger}$ ISPIC, Centre Tecnològic Telecomunicacions Catalunya, Barcelona, Spain. \\ Email: \{rpereira, xmestre\}@cttc.es \\
    $^{\ddagger}$ Software Radio Systems, Barcelona, Spain. Email: david.gregoratti@ieee.org
}

\maketitle

\begin{abstract}
Environmental scene reconstruction is of great interest for autonomous robotic applications, since an accurate representation of the environment is necessary to ensure safe interaction with robots. Equally important, it is also vital to ensure reliable communication between the robot and its controller. Large Intelligent Surface (LIS) is a technology that has been extensively studied due to its communication capabilities. Moreover, due to the number of antenna elements, these surfaces arise as a powerful solution to radio sensing. This paper presents a novel method to translate radio environmental maps obtained at the LIS to floor plans of the indoor environment built of scatterers spread along its area. The usage of a Least Squares (LS) based method, U-Net (UN) and conditional Generative Adversarial Networks (cGANs) were leveraged to perform this task. We show that the floor plan can be correctly reconstructed using both local and global measurements.  
\end{abstract}


\section{Introduction}
Mobile robots with mapping devices have been used to estimate indoor floor plans. The state-of-the-art methods are usually based on optical sensors to obtain the indoor maps in detail. For instance, \gls{lidar} \cite{surmann2003autonomous}, depth cameras \cite{zhang2012walk} and RGB cameras \cite{gao2014jigsaw}. Although these methods achieve acceptable accuracy, they have some limitations. \gls{lidar} might not be able to capture all types of materials meanwhile approaches based on cameras are dependent on the lighting conditions. To circumvent this, acoustic-based approaches such as microphones \cite{zhou2017batmapper} and ultrasonic sensors \cite{chong1999feature} are robust to lighting conditions. However, they have a limited sensing range and might malfunction in noisy environments.

On a related note, radar devices such as millimeter Wave (mmWave) radars have become popular for indoor sensing applications. They actively transmit \gls{rf} signals to monitor the reflections to sense nearby scatters' parameters such as range, speed or angle. Hence, they can be used indoors in poor lighting conditions. These radars have been used in applications such as human sensing \cite{singh2019radhar} and floor-plan reconstruction \cite{lu2020see}. However, mmWave radars work in high frequency bands, leading to short wavelengths. This leads to high energy attenuation over distance and weak penetrability through walls.

In the context of 6G, sensing has become a fundamental feature. 
For instance, the authors in \cite{weiss2021joint} explore this sensing-style capabilities of a mMIMO BS to jointly learn an antenna selection and a range-azimuth map of a beamforming gain. With the increasing number of receivers, \gls{lis} becomes a natural extension of the massive MIMO technology which designates a continuous electromagnetic surface able to transmit and receive radio waves. In practice, they are planar arrays conformed by a huge amount of closely spaced tiny antenna elements. There is a vast range of studies analyzing its application in communications~\cite{williams2021multiuser}. However, very little has been studied regarding its sensing capability~\cite{vaca2021assessing}. Consequently, in  \cite{vaca2021radio} we presented a method that enables reconstructing a radio map of the propagation environment using an indoor \gls{lis} deployment in the ceiling. This allows for tracking both active and passive users.


Motivated by these results and on the increasing interest in both sensing and \gls{lis}, in this work, we further explore the capabilities of \gls{lis} and provide a floor plan reconstruction based on the \gls{lis} received signal.  We leverage the usage of \gls{dgm} to learn a map from the complex-valued received signals at the \gls{lis} to the RGB image of the corresponding floor plan. Differently from previous works \cite{surmann2003autonomous, zhou2017batmapper, lu2020see}, we are mainly concerned with enabling sensing capabilities in a system primarily designed for communication. Consequently, such applications further exploit the usability of already deployed hardware. We assess the reconstruction performance with the original (ground-truth) floor plan composed of multiple elements representing, for example, robots, furniture and appliances 

\vspace{-0.5em}
\section{System and Problem Formulation}
\vspace{-0.5em}
Consider an indoor factory scenario in which $K_a$ active devices are deployed. These devices can be anything using a transmitter such as robots, smartphones or IoT devices. More importantly, since they are active devices, we can assume these devices to be communicating with a receiver through a wireless channel. In a real-world scenario, this communication often assists with the task at hand, e.g., sending commands to the robots. In our scenario, this receiver is a \gls{lis} placed on the factory's roof and, apart from assisting with the communication task, the \gls{lis} also tries to map the current environment of this factory.  This environment sensing can be divided into two steps:
\begin{enumerate}
    \item Mapping from the \textit{environmental signals} into a radio map of the environment from the \gls{lis} viewpoint, similar to \cite{vaca2021radio}. This allows converting the complex-valued data into a RGB real-valued data which is more natural to neural networks.
    \item Translation of the radio map into its current floor plan pattern, i.e., the disposition of the elements in this environment. In summary,  based on the raw complex-valued signals received at the \gls{lis}, we aim to reconstruct the current arrangement of the passive (non-active) elements present in the environment.
\end{enumerate}
For the second step, we rely on \gls{cgan} which has been widely used in the literature for the task of image-to-image reconstruction \cite{isola2017image}.  

Moreover, we assume this \gls{lis} to be equipped with $N=N_x \times N_y$ antenna elements and its physical aperture comprises its whole area. Concretely, we consider a square \gls{lis} composed of isotropic antennas and physical effects such as mutual coupling are ignored. A bit more formally, the sensing problem consists in obtaining a radio map that describes the environment from the superposition of the received signals from each of the $1 < k \leq  K_a$ users at every element of the \gls{lis}.  This map contains information on the $K_a$ active devices involved in the scenario as well as the $K_p$ passive objects/scatters. 

The superposed complex baseband signal received at the \gls{lis} is given by  
\begin{equation}
    \label{eq:RecSignal}
	\mathbf{y} = \sum_{k=1}^{K_a}\mathbf{h}_kx_k + \mathbf{n},
\end{equation}
with $x_k$ the transmitted (sensing) symbol from user $k$ (we consider $x_k=1$ without loss  of  generality), $\mathbf{h}_k\in\mathbb{C}^{N\times 1}$ the channel vector from a specific position of user $k$ to each antenna-element, and $\mathbf{n}\sim\mathcal{CN}_{N}(\mathbf{0},\sigma^2\mathbf{I}_N)$ the noise vector. To avoid frequency selectivity, we consider a narrowband transmission.

\begin{figure}[t]
    \centering
    \subfloat[Floor Plan\label{fig:exemplary_map}] {\includegraphics[width=0.4\columnwidth]{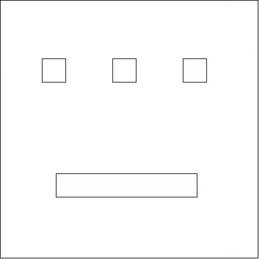}}
    \hspace{1em}
    \subfloat[$S=1$\label{fig:exemplary_map_1}]{\includegraphics[width=0.4\columnwidth]{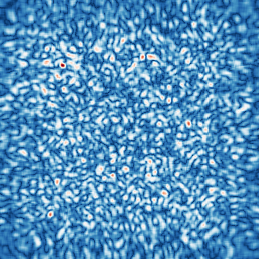}} 
    \hspace{1em} \\
    \subfloat[$S=100$\label{fig:exemplary_map_100}]{\includegraphics[width=0.4\columnwidth]{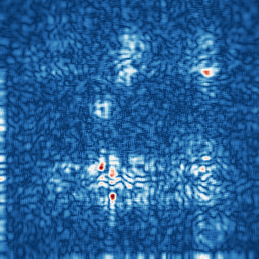}} 
    \hspace{1em}
    \subfloat[$S=1000$\label{fig:exemplary_map_1000}]{\includegraphics[width=0.4\columnwidth]{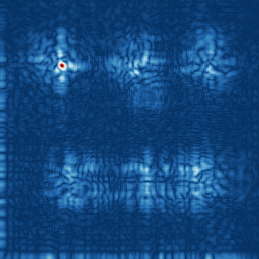}} 
    \caption{ Radio maps with its corresponding floor plan obtained over $S$-averaging channel samples acquired by \gls{lis} in a $\gamma=-10$ dB \gls{snr} condition.}
        \label{fig:exemplary_map:all}
\end{figure}

As a consequence of the large physical dimensions of the \gls{lis} in comparison with the distance from the transmitters to the roof, we need to account for spherical wave propagation during the modeling. The spherical-wave channel coefficient $h_{sp,n}$ at the LIS $n$-th element from an arbitrary active device transmission is proportional to \cite{zhou2015spherical}
\begin{equation} h_{sp,n} \ \propto\ \frac{1}{d_n}e^{-j\frac{2\pi}{\lambda}d_n},\label{eq:pattern} \end{equation}
where $d_n = \sqrt{(x_n - x_{k})^2 + (y_n - y_{k})^2 + (z_n - z_{k})^2}$ denotes the distance between the active device  $k$ and the $n$-th antenna. To obtain the radio map,  expressing the \gls{lis} in a vectorized version for ease of notation,  we can derive a \gls{mf} procedure such that \cite{vaca2021radio}
\begin{equation}
\label{eq:mf_map}
\mathbf{y}_{mf} =  \mathbf{h}_{sp} \ast \mathbf{y}, \end{equation}
denoting $\ast$ the spatial convolution operator, $\mathbf{h}_{sp} \in\mathbb{C}^{N_f\times 1}$ the expected spherical pattern (steering vector) for $N_f$ antennas \gls{lis} deployment on (\ref{eq:pattern}), $\mathbf{y}$ the received signal from (\ref{eq:RecSignal}) and $\mathbf{y}_{mf} \in\mathbb{C}^{N\times 1}$ the filtered output signal. As the convolution operator would reduce the output dimension (Due to the 2D convolution along the received signal at the \gls{lis}), we zero-pad $\mathbf{y}$ such that we guarantee $\mathbf{y}_{mf} \in \mathbb{C}^{N\times 1}$ dimension. To obtain a radio map, we just need to compute the energy at the output of the \gls{mf} procedure $\lvert \mathbf{y}_{mf}\rvert \in\mathbb{R}^{N \times 1}$. We then map the values to the RGB scale such that $F: \mathbb{R}^{N\times 1} \rightarrow  \{[0,255]\cap\mathbb{N} \}^ {N \times 3}$
\begin{equation}
    \mathbf{y}_m = F(\lvert \mathbf{y}_{mf}\rvert) \label{eq:r_map}
\end{equation}
 that represents the radio map. Please note, we have to know the frequency $f$ and assume a distance $z$ from the transmitter to the \gls{lis} to design the filter\footnote{For a more detailed explanation of the radio maps, we gently refer the readers to \cite{vaca2021radio}. The distance $z$ is a parameter for the filter design. This does not imply that in the evaluation, all the transmitters or scatters  are fixed at this distance.}. In our work, we design a filter for $f=3.5$ GHz, $z=8$ m and $N_f=100\times100$ antenna elements $\frac{\lambda}{2}$ spaced\footnote{Note we use a fixed kernel with $z=8$ m which corresponds to the height of the building. We do not need to calibrate the \gls{mf} to specific distances.}. 
Let us assume we can obtain $S$ samples at each channel coherence interval. In this way, we can use these extra samples to perform an $S$-averaging of the received signal measurements at the \gls{lis} viewpoint, reducing the noise variance contribution and enhancing the quality of the obtained radio map. Figure~\ref{fig:exemplary_map:all}(a) shows a ground-truth floor plan. Looking at Figure~\ref{fig:exemplary_map:all}(b)-(d)  we can see three exemplary radio maps representing the ground-truth floor plan. The radio map captures the reflections of the scatters (rectangular and square shapes representing walls/objects) that act as virtual sources. The $S$-averaging effect in reducing the noise contribution leads to an enhancement in the radio map quality. The target of our model will be translating these radio maps to the ground-truth floor plan.

\subsection{Received signal and noise modeling}
In order to simulate the propagation environment in the most reliable way, we resort to ray tracing\cite{FEKO}. From the ray-tracing simulation, the received signal in (\ref{eq:RecSignal}) is obtained as the complex electric field arriving at the $n$-th antenna element, $\widetilde{E}_{n}$, which can be regarded as the superposition of each ray path $r \in N_r$ from every $k \in K_a$ user, i.e., 
\begin{equation}
    \label{eq:Esum}
  \widetilde{E}_{n} = \sum_{k=1}^{K_a}\sum_{r=1}^{N_r} \widetilde{E}_{n,r,k}= \sum_{k=1}^{K_a}\sum_{r=1}^{N_r}E_{n,r,k} e^{j\phi_{n,r,k}}.  
\end{equation}
Then, the complex signal at the output of the $n$-th element is therefore given by 
\begin{equation}
    \label{eq:complexSignal}
    y_n = \sqrt{\frac{\lambda^2Z_n}{4\pi Z_0}} \widetilde{E}_{n} + n_n,
\end{equation}
with $\lambda$ the wavelength, $Z_0 = 120\pi$ the free space impedance and  $Z_n$ the antenna impedance. 
For simplicity, we consider $Z_n = 1\,\forall\, n$. Finally, we define the \gls{snr}, $\gamma$, as
\begin{equation}
    \label{eq:snr}
    \gamma \triangleq  \frac{\lambda^2}{4\pi Z_0 N \sigma^2}\displaystyle\sum_{n=1}^{N} |\widetilde{E}_{n}|^2,
\end{equation}
where $N$ denotes the number of antenna elements in the \gls{lis}.


\vspace{-0.5em}
\section{Reconstruction learning}
\label{sec:reconstruction_learning}
\vspace{-0.5em}
\subsection{Least Squares}
This method tries to naively find the best linear mapping $\mathbf{W}^* \in\mathbb{R}^{N\times N}$ from the $i$-th radio map $\mathbf{y}_{m}^{(i)} \in\mathbb{R}^{N}$ into the $i$th floor plan $\mathbf{x}^{(i)} \in\mathbb{R}^{N}$ by minimizing the average least square error 
\begin{equation}
\mathbf{W}^* = \arg \min_\mathbf{W} \frac{1}{\mathrm{T}}\sum_{i = 1}^{\mathrm{T}} \left(\mathbf{x}^{(i)} - \mathbf{W}\mathbf{y}_m^{(i)}\right)^2
\end{equation}
over the $T$ training samples. Prediction is then performed on a new sample by $\mathbf{\hat{x}}^{(j)} = \mathbf{W}\mathbf{y}_m^{(j)}$. Despite its simplicity, similar solutions have been widely applied in many signal-to-image reconstruction, such as
 medical applications~\cite{5446399} and  synthetic aperture radars sensing~\cite{kang2019compressive}. 
  In our scenario, LS method often returns a noisy version of the environment reconstruction. Hence, to simulate the best possible set of post-processing operations that improve the LS estimator,  we redefine $\mathbf{\hat{x}}^{(j)} = \min(\mathbf{x}^{(j)}, \hat{\mathbf{x}}^{(j)})$ with the entry-wise minimum operator, i.e., we remove the noise area that is present outside of the area of interest.

\vspace{-1em}
\subsection{U-Net}
U-nets are essentially autoencoder networks with skipped connections \cite{ronneberger2015u}. At the encoder side, we learn the feature mapping of an image while converting it to a vector. U-Net extends this vector to a segmented picture, utilizing the same feature maps as were used for compression at the encoder side (i.e. skipped connections). This would keep the image's structural integrity and provide information to the decoder to perform the segmentation. 
Image segmentation can be applied to our specific problem. We can see the groundtruth floor plan $\mathbf{x}$ as the segmentation of the radio map $\mathbf{y}_m$ composed of two classes, background and shapes. To reduce training time and prevent overfitting, we will use a pre-trained model for the encoder, the mobileNet-v2 \cite{sandler2018mobilenetv2}, while for the decoder we will use the up-sampling block of Pix2Pix \cite{isola2017image}. Finally, the U-Net minimizes the Binary Cross-Entropy (BCE) loss. 



\subsection{Conditional Generative Adversarial Network}

We design our problem as a cGANs learning procedure by designing a generator for the distribution $p_g$ over the floor plan data of the environment $\mathbf{x}$, given as conditional information the radio map $\mathbf{y}_m$ from (\ref{eq:r_map}). The generator models a mapping function from a prior noise distribution $p_{\mathbf{z}}(\mathbf{z})$ to the floor plan space $G(\mathbf{z}|\mathbf{y}_m;\theta_g)$. Similarly, we design a discriminator $D(\mathbf{x}|\mathbf{y}_m; \theta_d)$ that outputs a scalar value representing the probability that $\mathbf{x}$ came form training data instead of $p_g$. 
The target min-max cost function is given by
\begin{align}
\label{eq:cGANs_cf}
\mathcal{L}_{cGANs} = \min_{G}\max_{D}\mathbb{E}_{\mathbf{x}\sim p_{\text{data}}(\mathbf{x})}[\log{D(\mathbf{x}|\mathbf{y}_m)}] + \nonumber \\ \mathbb{E}_{\mathbf{z}\sim p_{\mathbf{z}}(\mathbf{z})}[1 - \log{D(G(\mathbf{z}|\mathbf{y}_m))}].
\end{align} 

\begin{figure*}[!ht]
    \captionsetup[subfigure]{labelformat=empty}
    \centering
    \subfloat[(a) Scenario 1 - Original] {\includegraphics[width=0.4\columnwidth]{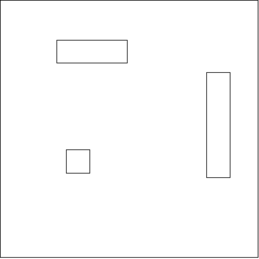}}
    \hspace{1em}
    \subfloat[(c) Scenario 1 - LS ]{\includegraphics[width=0.4\columnwidth]{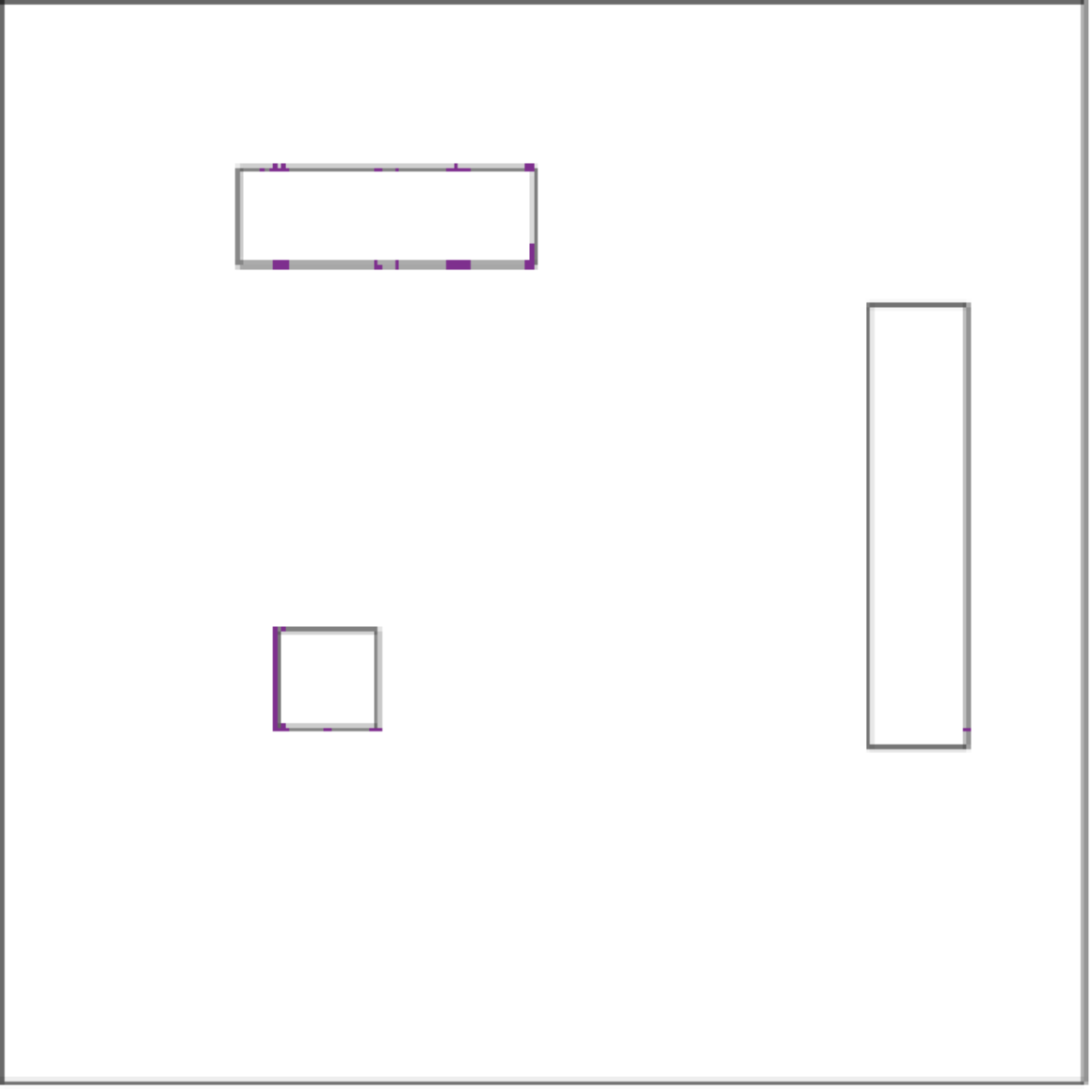}} 
    \hspace{1em} 
    \subfloat[(e) Scenario 1 -  U-Net] {\includegraphics[width=0.4\columnwidth]{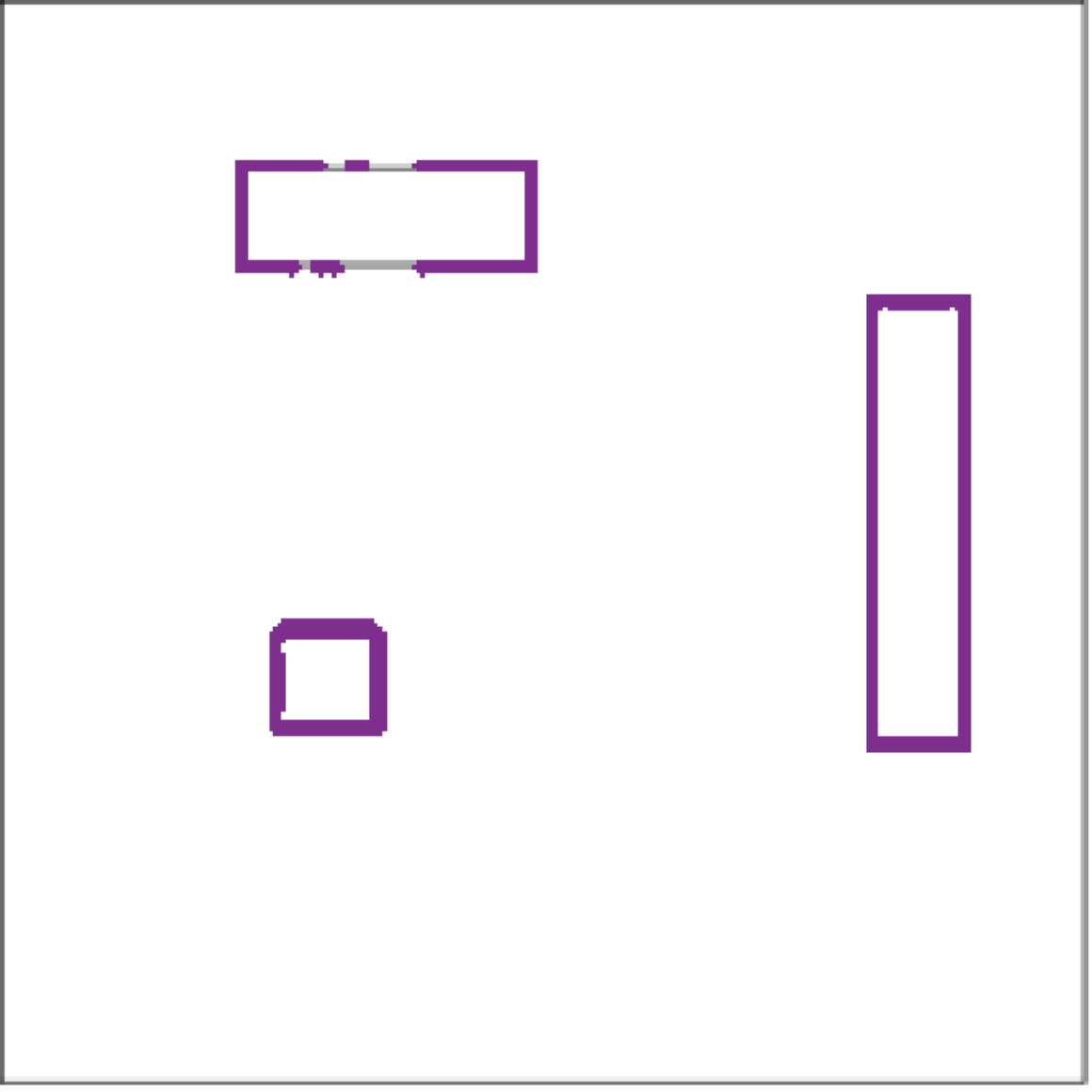}} 
    \hspace{1em}
    \subfloat[(g) Scenario 1 - cGANs]{\includegraphics[width=0.4\columnwidth]{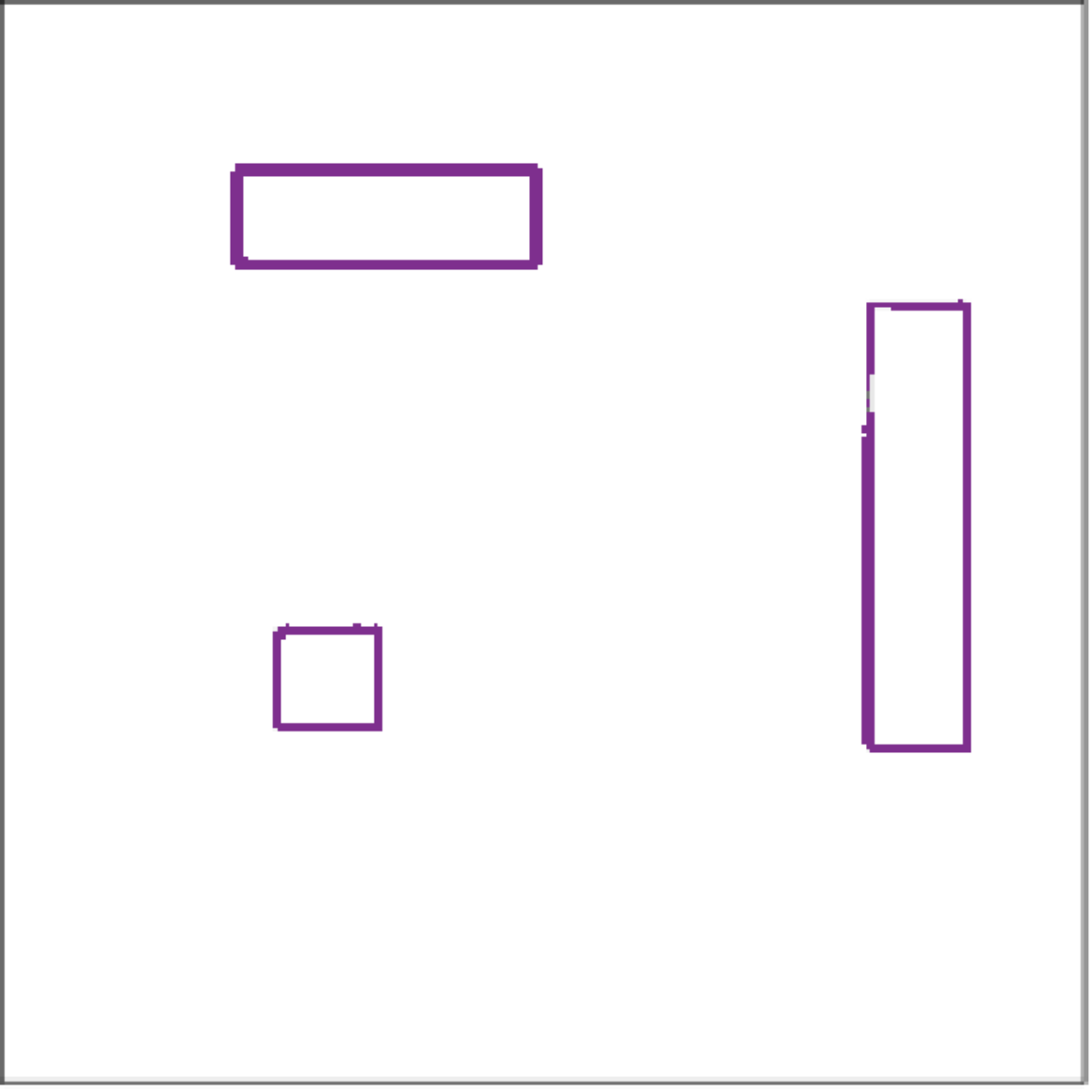}} 
    \vspace{1em} \hspace{1em} \\
    \subfloat[(b) Scenario 2 - Original ]{\includegraphics[width=0.4\columnwidth]{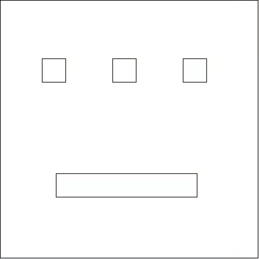}} 
    \hspace{1em}
    \subfloat[(d) Scenario 2 - LS  ]{\includegraphics[width=0.4\columnwidth]{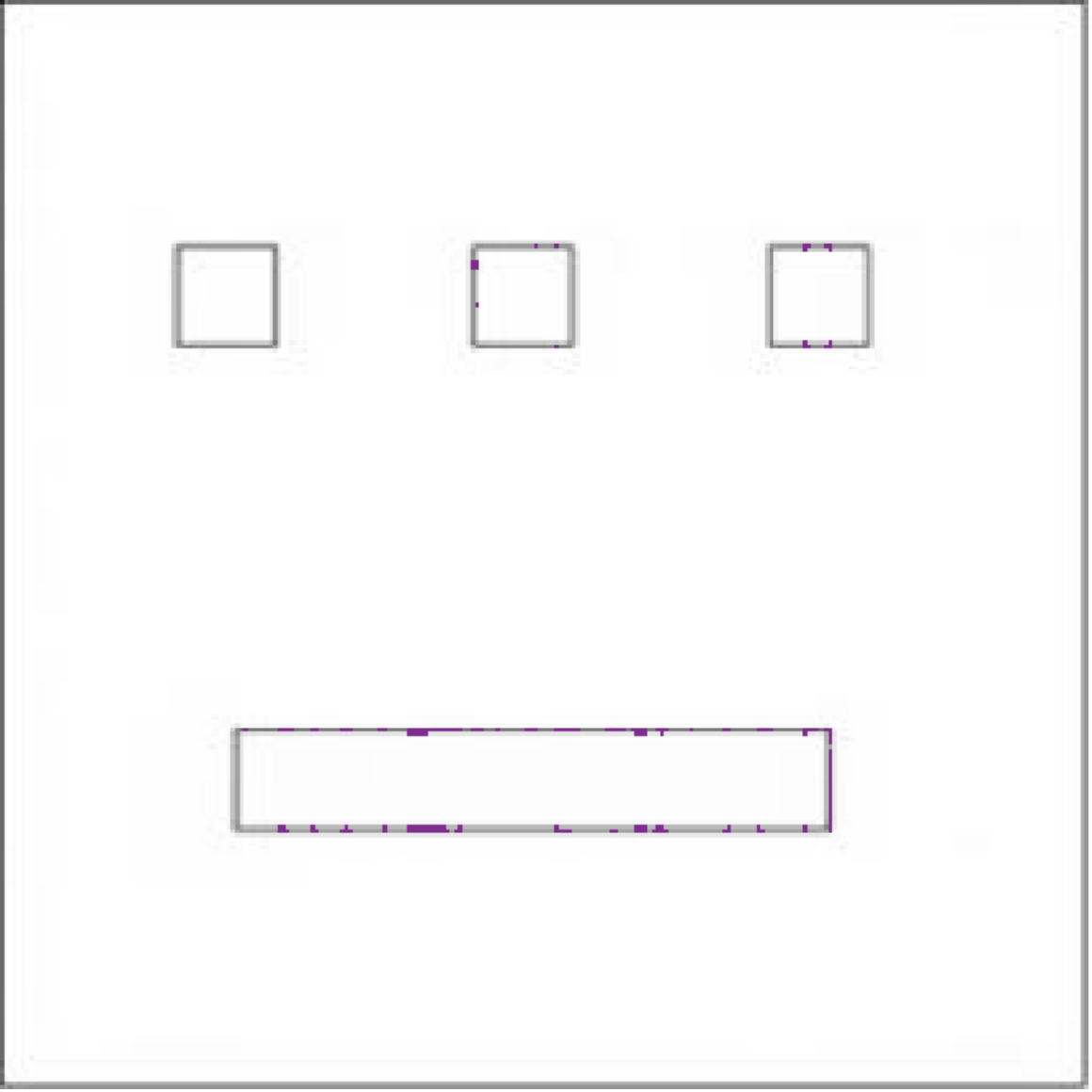}} 
    \hspace{1em} 
    \subfloat[(f) Scenario 2 - U-Net ]{\includegraphics[width=0.4\columnwidth]{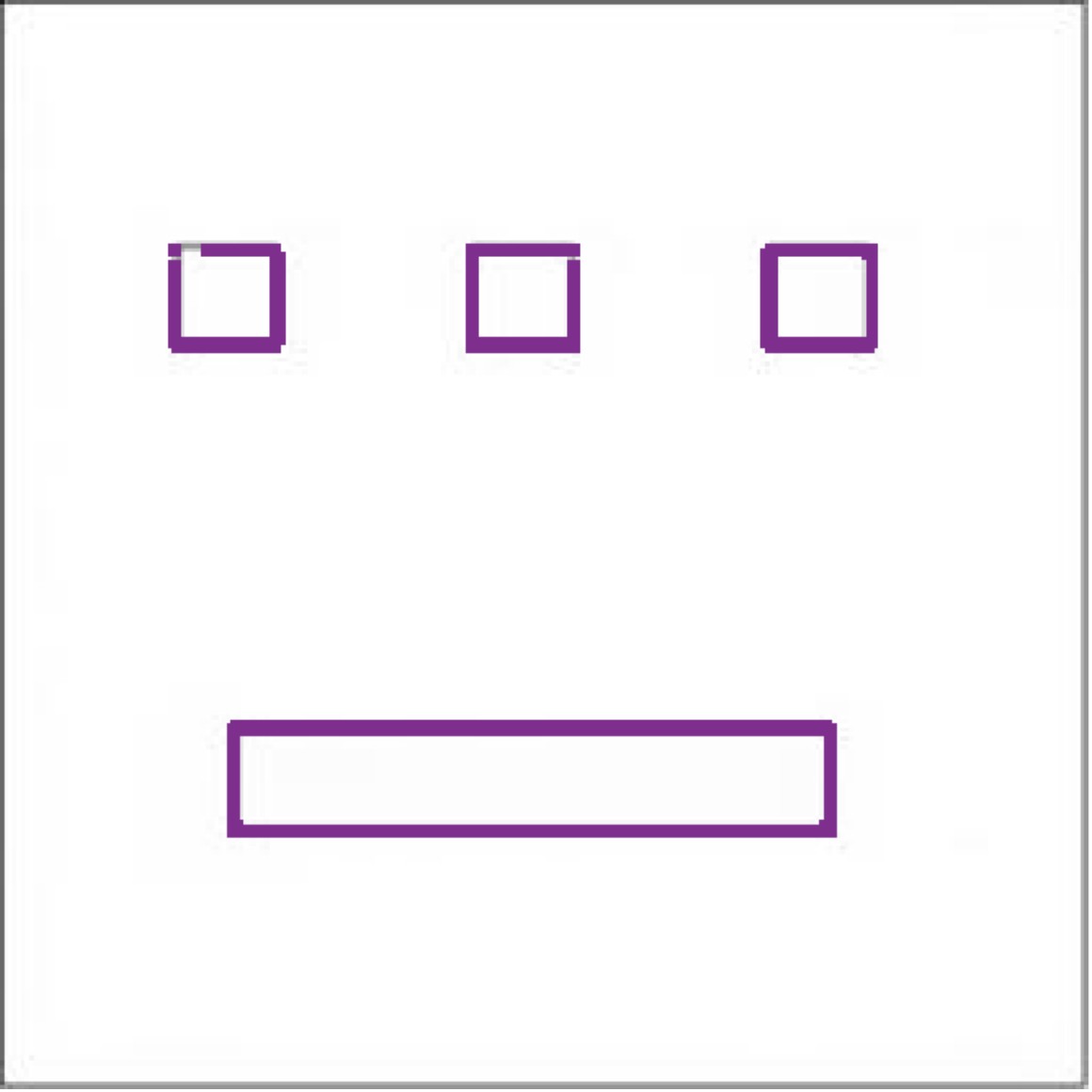}} 
    \hspace{1em}
    \subfloat[(h) Scenario 2 - cGANs]{\includegraphics[width=0.4\columnwidth]{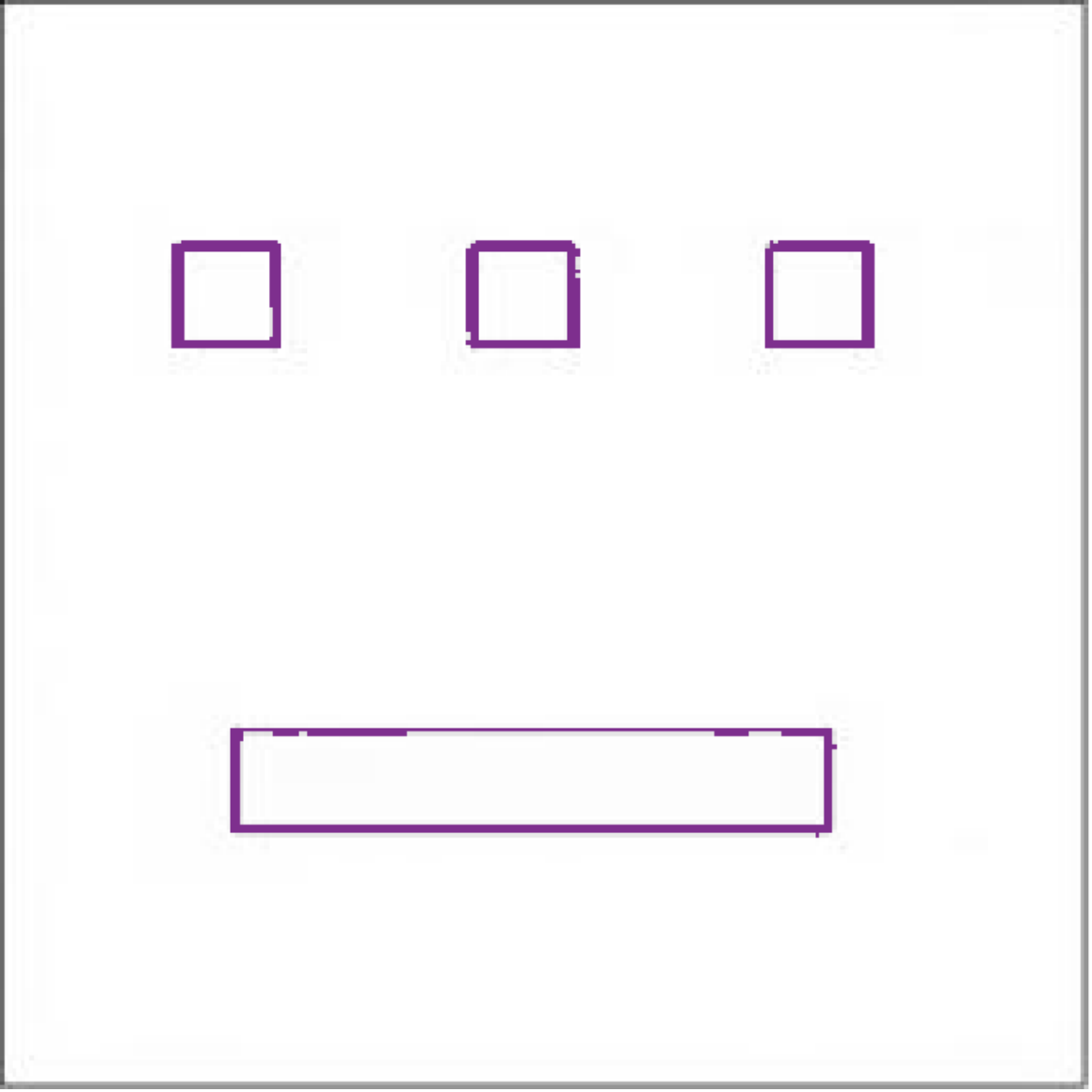}} 
    \caption{Visual representation of the two original (a)-(b) floor plans and their and reconstructions (c)-(h) using the three different methods. Reconstructions are highlighted in purple and the original floor plans in black. }
   \label{fig:scenarios_reconstruction}
\end{figure*}

More specifically, we re-implement a well-known model used in the field of image-to-image translation \cite{wang2018high}. The main difference with respect to the general \gls{cgan} framework resides in the loss function.  The authors of \cite{wang2018high} improve the adversarial loss by incorporating a feature matching loss. It works by extracting features from several layers of the discriminator, trying to match these in-between representations among the real and the fake data. The new loss function can be expressed as
\begin{align}
\label{eq:cGANs_pix2pix}
\mathcal{L}^*_{cGANs} = \min_{G}\max_{D} \mathcal{L}_{cGANs} 
\\ + \nonumber  \lambda\mathbb{E}_{\mathbf{x}, \mathbf{y}_m, \mathbf{z}}\sum_{l=1}^L\frac{1}{N_l} [\|D^{(l)}(\mathbf{x}|\mathbf{y}_m) - D^{(l)}(\mathbf{z}|\mathbf{y}_m)\|_1],
\end{align}
where $L$ denotes the number of layers, $N_l$ denotes the number of elements in each layer and $\lambda$ is an hyper-parameter that controls the weight of the terms. The model will learn a mapping between the wireless environmental signals and its corresponding floor plan, i.e., it will perform a translation from the radio map $\mathbf{y}_m$ to the floor plan $\mathbf{x}$.


\vspace{-0.5em}
\section{Dataset description}
\vspace{-0.5em}
In the ray-tracing simulation, we consider two scenarios of size $10.34 \times 10.34 \times 8$ m with different scatters arrangement. We deploy an \gls{lis} with $259\times259$ elements separated $\lambda/2$ apart. Every $K_a$ active device transmits a narrowband signal of 20 dBm at $3.5$ GHz. The distance from which the \gls{mf} is calibrated is $z=8$ m (building height). The scatters  are modeled as metallic (with conductivity $s=19444$ S/m, relative permittivity $\epsilon=1$ and relative permeability $\mu=20$) and the walls as brick (with $s=0.078$ S/m, $\epsilon=4$ and $\mu=1$)\footnote{These values are provided by the software manual \cite{FEKO}.}. Figure~\ref{fig:scenarios_reconstruction}(a)-(b) show the distribution of the metallic elements in the two floor plans considered in this work. The dataset 
is obtained by sampling the received signal at each element of the \gls{lis} from all the different active $K_a$ in the scenario. Then the resulting signal is processed following the process described in Section II to generate the radio map. 

To guarantee generalization, we conform a dataset composed by a selection of $S  \in [1, 100, 1000]$ and $K_a \in [5, 20]$ for the two scenarios presented. Finally, we are primarily interested in poor signaling conditions, hence, we set $\gamma = -10$ dB which means that there is much noise in the communication between the active transmitters and the \gls{lis}.  Then, formally, the dataset is denoted as $\{\mathbf{y}_m^{(i)}, \mathbf{x}^{(i)}\}_{i=1}^T$, where $\mathbf{y}_m^{(i)}$ is the $i$-{th} $N$-dimensional training input features vector (radio map) and $\mathbf{x}^{(i)}$ is the $i$-{th} $N$-dimensional target vector (floor plan). Our dataset is composed of $2400$ samples which are split into 70/10/20 for training, validation and test set, respectively.

\bgroup
\def\arraystretch{1.2}
\begin{table*}[h]
\centering
\begin{tabular}{c|c|c|c|ccc|}
\cline{2-7}
\multicolumn{1}{l|}{}                     & \multirow{2}{*}{{Method}} & \multirow{2}{*}{{PSNR}} & \multirow{2}{*}{{SSIM}} & \textbf{}                             & {Error distance (cm)}                     & \textbf{}       \\ \cline{5-7} 
                                          &                                  &                                &                                & \multicolumn{1}{c|}{\textbf{$K_a = 5$}}  & \multicolumn{1}{c|}{\textbf{$K_a = 20$}} & {Average}    \\ \hline
\multicolumn{1}{|c|}{}                    & LS                               & $12.71 \pm 1.29$               & $0.19 \pm 0.03$                & \multicolumn{1}{c|}{-}                & \multicolumn{1}{c|}{-}                & -               \\ \cline{2-7} 
\multicolumn{1}{|c|}{\textit{Scenario 1}} & U-Net                            & $21.51 \pm 0.24$               & $0.93$                         & \multicolumn{1}{c|}{$5.68 \pm 5.67 $} & \multicolumn{1}{c|}{$3.51 \pm 3.50$}  & $4.59\pm4.59$   \\ \cline{2-7} 
\multicolumn{1}{|c|}{}                    & cGANs                             & $\mathbf{31.77 \pm 0.3}$                & $\mathbf{0.95}$                         & \multicolumn{1}{c|}{$\mathbf{4.58 \pm 4.50}$}  & \multicolumn{1}{c|}{$\mathbf{1.96  \pm 1.95}$} & $\mathbf{3.27 \pm 3.27}$ \\ \hline
\multicolumn{1}{|c|}{}                    & LS                               & $11.896 \pm 1.09$              & $0.232 \pm 0.04$               & \multicolumn{1}{c|}{-}                & \multicolumn{1}{c|}{-}                & -               \\ \cline{2-7} 
\multicolumn{1}{|c|}{\textit{Scenario 2}} & U-Net                            & $21.51 \pm 0.17$               & $0.94$                         & \multicolumn{1}{c|}{$\mathbf{5.27}$}           & \multicolumn{1}{c|}{$\mathbf{3.64  \pm 8.02}$} & $\mathbf{4.45 \pm 4.01}$ \\ \cline{2-7} 
\multicolumn{1}{|c|}{}                    & cGANs                             & $\mathbf{32.66 \pm 0.24}$               & $\mathbf{0.95}$                         & \multicolumn{1}{c|}{$7.99 + \pm 8.9$} & \multicolumn{1}{c|}{$7.89$}           & $7.94 \pm 4.45$ \\ \hline
\end{tabular}
\caption{\label{tab:avg_comparison_reconstructions} Average qualitative comparison of LS, U-Net and cGANs in terms of PSNR, SSIM and scattering centroids error distance.}
\end{table*}
\egroup

\begin{figure}[h]
    \centering
    \definecolor{mycolor1}{rgb}{0.85000,0.32500,0.09800}%
\definecolor{mycolor2}{rgb}{0.49400,0.18400,0.55600}%
\definecolor{mycolor3}{rgb}{0.46600,0.67400,0.18800}%
\begin{tikzpicture}
\begin{axis}[%
width=0.8\columnwidth,
height=0.5\linewidth,
at={(1.011in,1.2in)},
scale only axis,
bar shift auto,
xmin=0.0,
xmax=7.5,
xtick={1,2,3,4.5,5.5,6.5},
xticklabels={{1}, {100}, {1000}, {1}, {100}, {1000}},
ymin=0,
ymax=45,
ylabel={PSNR},
axis background/.style={fill=white},
legend style={legend cell align=center, align=center, draw=white!15!black, legend style={at={(1.35in, 0.533\linewidth)}, anchor=center,legend columns=3, legend},
}
]
\addplot[ybar, bar width=0.178, fill=mycolor1, draw=black, area legend] table[row sep=crcr] {%
1	14.255\\
2	12.545\\
3	11.325\\
4.5	14.145\\
5.5	12.375\\
6.5	11.12\\
};
\addlegendentry{LS }
\addplot[ybar, bar width=0.178, fill=mycolor2, draw=black, area legend] table[row sep=crcr] {%
1	21.02\\
2	21.765\\
3	21.745\\
4.5	22.05\\
5.5	21.98\\
6.6	21.985\\
};

\addlegendentry{U-Net }
\addplot[ybar, bar width=0.178, fill=mycolor3, draw=black, area legend] table[row sep=crcr] {%
1	24.265\\
2	35.475\\
3	35.585\\
4.5	24.165\\
5.5	36.365\\
6.5	36.17\\
};
\addlegendentry{cGANs }

\draw[color=gray,  thick] (0.4,0) rectangle (3.6,38);

\filldraw[darkgray] (1, 40) circle (0pt) node[anchor=west]{\shortstack{\textit{Scenario 1}}};

\filldraw[darkgray] (4.5, 40) circle (0pt) node[anchor=west]{\shortstack{\textit{Scenario 2}}};

\draw[color=gray,  thick] (3.9,0) rectangle (7.2,38);

\end{axis}
\end{tikzpicture}%
    \caption{Comparison of floor plan reconstruction quality based on the Average \gls{psnr} metric along the test set for varying number of $S$. }
   \label{fig:reconstruction_quality}
   \vspace{-1.4em}
\end{figure}
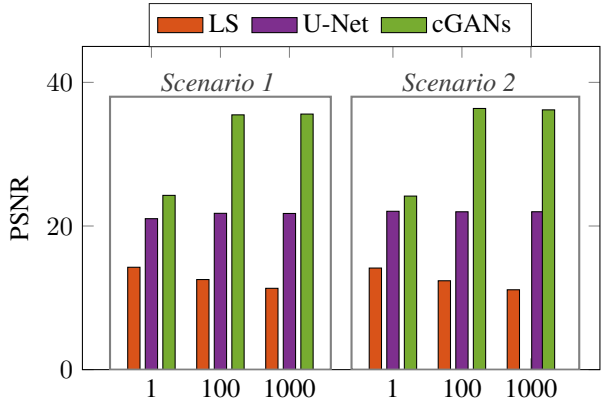

\vspace{-0.5em}
\section{Numerical results and Discussion}
\vspace{-0.5em}

In order to validate the proposed methodology,   we rely on standard image quality metrics, to evaluate the performance of the reconstructed environment using the three methods considered in this work (LS, U-Net and cGANs), specifically, we evaluate these methods using  \gls{psnr} and \gls{ssim} \cite{hore2010image} as quality metrics. They provide a general evaluation of how similar the reconstruction is to the original floor plan. However, this evaluation is performed based on the whole image. Hence, we also calculate the distance (in centimeters) between the centres of the scatters in the original and predicted environment. This further provides us with a way to assess local spatial predictions.

Let us first analyse the general reconstructions capability of each of these methods. Figure~\ref{fig:reconstruction_quality} compares the average PSNR obtained on the test set by the three methods considering different $S$-averaging of the received signal. As expected, the original signal $S=1$ provides little information and hence makes it unfeasible to obtain good quality reconstruction (perfect reconstruction would lead to a maximum PSNR of $48$ dB). Moreover, we notice that specifically for the linear mapping (LS), the quality of the reconstruction drastically decays as $S$-averaging increases. This happens despite the post-processing mechanisms performed (see Sec. \ref{sec:reconstruction_learning}) and the quality of the data used for training - we also considered training different models for specific scenarios, i.e., $S$-averaging and  number of active transmitters, but obtained similar results. Consequently, it is reasonable to assume that it is extremely hard to find a direct linear mapping from the input radio map to the desired output. This further motivates the application of machine learning to this problem. In fact, for U-Net and cGANs, we notice the opposite behaviour, i.e., the larger $S$, the larger the PSNR obtained. In all scenarios the best results are obtained using cGANs. 
The overall generalisation of the methods is compared by considering the average reconstruction results over all different signal configurations $S \in [1, 100, 1000]$, $K_a \in [5, 20]$ and for the two different scenarios. 
Table~\ref{tab:avg_comparison_reconstructions} contains the PSNR, SSIM and average distance error of the predicted scattered objects\footnote{We do not denote the variance of the results when it is negligible.}. The best results are highlighted.   Indeed, the average quantity of these metrics is consistent with what is explained above. More interesting, though, is the error distance (in centimetres) between the central position of the scattered objects and their reconstruction. For the original floor plan, these centres are the middle point of each rectangle. In the reconstructed images, we obtain these central positions by approximating polygons to the scatter. The cGANs exhibit the best performance in most of the comparative analyses. It obtains the highest PSNRs and SSIM in both scenarios and the smallest average distance for \textit{Scenario 1}.  Specifically, for \textit{Scenario 2}, U-Net obtains smaller error distances among the true and the predicted position of the scatters. However, in real-world space, this difference is almost negligible.  An exemplary visual reconstruction for $S = 100$ is provided in Figure~\ref{fig:scenarios_reconstruction}(e)-(h) for both U-Net and cGANs. Finally, since LS provides little information (see Figure~\ref{fig:scenarios_reconstruction}(c)-(d) for a visual comparison) on the reconstructed scenario, it is impossible to perform any analysis on the error distance between the true location of the scatters and their predicted locations.

\vspace{-0.5em}
\section{Conclusion}
\vspace{-0.5em}
We have demonstrated a proof-of-concept that it is possible to learn an environmental scene reconstruction using signals received at a \gls{lis} while it is performing communication. This sensing procedure can be done in parallel and without interfering with the communication task. We have shown that using both U-Net and cGANs it is possible to accurately (with less than 16 cm error) estimate the central position of the scatters present in the environment directly from these image reconstructions. In future work, we plan to explore real-world environments as well as explore the learning environment to enhance communication performance.

\bibliographystyle{IEEEtran}
\bibliography{./biblio.bib}

\begin{thebibliography}{10}
\providecommand{\url}[1]{#1}
\csname url@samestyle\endcsname
\providecommand{\newblock}{\relax}
\providecommand{\bibinfo}[2]{#2}
\providecommand{\BIBentrySTDinterwordspacing}{\spaceskip=0pt\relax}
\providecommand{\BIBentryALTinterwordstretchfactor}{4}
\providecommand{\BIBentryALTinterwordspacing}{\spaceskip=\fontdimen2\font plus
\BIBentryALTinterwordstretchfactor\fontdimen3\font minus
  \fontdimen4\font\relax}
\providecommand{\BIBforeignlanguage}[2]{{%
\expandafter\ifx\csname l@#1\endcsname\relax
\typeout{** WARNING: IEEEtran.bst: No hyphenation pattern has been}%
\typeout{** loaded for the language `#1'. Using the pattern for}%
\typeout{** the default language instead.}%
\else
\language=\csname l@#1\endcsname
\fi
#2}}
\providecommand{\BIBdecl}{\relax}
\BIBdecl

\bibitem{surmann2003autonomous}
H.~Surmann, A.~N{\"u}chter, and J.~Hertzberg, ``{An autonomous mobile robot
  with a 3D laser range finder for 3D exploration and digitalization of indoor
  environments},'' \emph{Robotics and Autonomous Systems}, vol.~45, no. 3-4,
  pp. 181--198, 2003.

\bibitem{zhang2012walk}
Y.~Zhang, C.~Luo, and J.~Liu, ``{Walk\&Sketch: Create Floor Plans with an RGB-D
  Camera},'' in \emph{Proceedings of the 2012 ACM Conference on Ubiquitous
  Computing}, 2012, pp. 461--470.

\bibitem{gao2014jigsaw}
R.~Gao, M.~Zhao, T.~Ye, F.~Ye, Y.~Wang, K.~Bian, T.~Wang, and X.~Li, ``{Indoor
  Floor Plan Reconstruction via Mobile Crowdsensing},'' in \emph{Proceedings of
  the 20th annual international conference on Mobile computing and networking},
  2014, pp. 249--260.

\bibitem{zhou2017batmapper}
B.~Zhou, M.~Elbadry, R.~Gao, and F.~Ye, ``{BatMapper: Acoustic Sensing Based
  Indoor Floor Plan Construction Using Smartphones},'' in \emph{Proceedings of
  the 15th Annual International Conference on Mobile Systems, Applications, and
  Services}, 2017, pp. 42--55.

\bibitem{chong1999feature}
K.~S. Chong and L.~Kleeman, ``{Feature-based Mapping in Real, Large Scale
  Environments using an Ultrasonic Array},'' \emph{The International Journal of
  Robotics Research}, vol.~18, no.~1, pp. 3--19, 1999.

\bibitem{singh2019radhar}
A.~D. Singh, S.~S. Sandha, L.~Garcia, and M.~Srivastava, ``{RadHAR: Human
  Activity Recognition from Point Clouds Generated through a Millimeter-wave
  Radar},'' in \emph{Proceedings of the 3rd ACM Workshop on Millimeter-wave
  Networks and Sensing Systems}, 2019, pp. 51--56.

\bibitem{lu2020see}
C.~X. Lu, S.~Rosa, P.~Zhao, B.~Wang, C.~Chen, J.~A. Stankovic, N.~Trigoni, and
  A.~Markham, ``{See Through Smoke: Robust Indoor Mapping with Low-cost mmWave
  Radar},'' in \emph{Proceedings of the 18th International Conference on Mobile
  Systems, Applications, and Services}, 2020, pp. 14--27.

\bibitem{weiss2021joint}
T.~Weiss, N.~Peretz, S.~Vedula, A.~Feuer, and A.~Bronstein, ``Joint
  optimization of system design and reconstruction in mimo radar imaging,'' in
  \emph{2021 IEEE 31st International Workshop on Machine Learning for Signal
  Processing (MLSP)}.\hskip 1em plus 0.5em minus 0.4em\relax IEEE, 2021, pp.
  1--6.

\bibitem{williams2021multiuser}
R.~J. Williams, P.~Ram{\'\i}rez-Espinosa, E.~De~Carvalho, and T.~L. Marzetta,
  ``Multiuser mimo with large intelligent surfaces: Communication model and
  transmit design,'' in \emph{ICC 2021-IEEE International Conference on
  Communications}.\hskip 1em plus 0.5em minus 0.4em\relax IEEE, 2021, pp. 1--6.

\bibitem{vaca2021assessing}
C.~J. Vaca-Rubio, P.~Ramirez-Espinosa, K.~Kansanen, Z.-H. Tan, E.~De~Carvalho,
  and P.~Popovski, ``Assessing wireless sensing potential with large
  intelligent surfaces,'' \emph{IEEE Open Journal of the Communications
  Society}, vol.~2, pp. 934--947, 2021.

\bibitem{vaca2021radio}
C.~J. Vaca-Rubio, P.~Ramirez-Espinosa, K.~Kansanen, Z.-H. Tan, and
  E.~de~Carvalho, ``{Radio Sensing with Large Intelligent Surface for 6G},''
  \emph{arXiv preprint arXiv:2111.02783}, 2021.

\bibitem{isola2017image}
P.~Isola, J.-Y. Zhu, T.~Zhou, and A.~A. Efros, ``{Image-to-Image Translation
  with Conditional Adversarial Networks},'' in \emph{Proceedings of the IEEE
  Conference on Computer Vision and Pattern Recognition}, 2017.

\bibitem{zhou2015spherical}
Z.~Zhou, X.~Gao, J.~Fang, and Z.~Chen, ``Spherical wave channel and analysis
  for large linear array in {LoS} conditions,'' in \emph{2015 IEEE Globecom
  Workshops (GC Wkshps)}.\hskip 1em plus 0.5em minus 0.4em\relax IEEE, 2015,
  pp. 1--6.

\bibitem{FEKO}
Feko, altair engineering, inc. \url{ https://www.altairhyperworks.com/feko}.

\bibitem{5446399}
N.~Vaswani, ``{LS-CS-Residual (LS-CS): Compressive Sensing on Least Squares
  Residual},'' \emph{IEEE Transactions on Signal Processing}, vol.~58, no.~8,
  2010.

\bibitem{kang2019compressive}
M.-S. Kang and K.-T. Kim, ``{Compressive sensing based SAR imaging and
  autofocus using improved Tikhonov regularization},'' \emph{IEEE Sensors
  Journal}, vol.~19, no.~14, pp. 5529--5540, 2019.

\bibitem{ronneberger2015u}
O.~Ronneberger, P.~Fischer, and T.~Brox, ``U-net: Convolutional networks for
  biomedical image segmentation,'' in \emph{International Conference on Medical
  image computing and computer-assisted intervention}.\hskip 1em plus 0.5em
  minus 0.4em\relax Springer, 2015, pp. 234--241.

\bibitem{sandler2018mobilenetv2}
M.~Sandler, A.~Howard, M.~Zhu, A.~Zhmoginov, and L.-C. Chen, ``Mobilenetv2:
  Inverted residuals and linear bottlenecks,'' in \emph{Proceedings of the IEEE
  Conference on Computer Vision and Pattern Recognition}, 2018, pp. 4510--4520.

\bibitem{wang2018high}
T.-C. Wang, M.-Y. Liu, J.-Y. Zhu, A.~Tao, J.~Kautz, and B.~Catanzaro,
  ``High-resolution image synthesis and semantic manipulation with conditional
  gans,'' in \emph{Proceedings of the IEEE conference on computer vision and
  pattern recognition}, 2018, pp. 8798--8807.

\bibitem{hore2010image}
A.~Hore and D.~Ziou, ``{Image quality metrics: PSNR vs. SSIM},'' in \emph{2010
  20th international conference on pattern recognition}.\hskip 1em plus 0.5em
  minus 0.4em\relax IEEE, 2010, pp. 2366--2369.

\end{thebibliography}

\end{document}